\documentclass{article}
\pdfoutput=1

\usepackage{graphicx}
\usepackage[pdftex,
            colorlinks=true,
            linkcolor=blue,
            citecolor=blue,
            urlcolor=blue,
            hyperindex={true}
            ]{hyperref}
\usepackage{amsmath,amssymb,bm}
\usepackage{color}
\usepackage{verbatim}
\usepackage[left=1.25in,top=1.25in,right=1.25in,bottom=1.25in,head=1.25in]{geometry}
\usepackage{amsfonts,amsmath,amssymb,amsthm}
\usepackage{verbatim,float,url,enumerate}
\usepackage{graphicx,subfigure,epsfig,psfrag}
\usepackage{dsfont,bm,color,appendix}
\newcommand{\be}{\begin{equation}}
\newcommand{\ee}{\end{equation}}

\newcommand{\mcalR}{\mathcal{R}}
\renewcommand{\vec}[1]{\bm{#1}}

\newcommand{\tx}{\vec{\theta_x}}
\newcommand{\tz}{\vec{\theta_z}}
\newcommand{\bxz}{b_{xz}}
\newcommand{\bxy}{b_{xy}}
\newcommand{\bzy}{b_{zy}}

\newcommand{\ltx}{\lambda^x_2}

\newcommand{\ltz}{\lambda^z_2}
\newcommand{\lx}{\lambda_x}
\newcommand{\lz}{\lambda_z}
\newcommand{\y}{\vec{y}}
\newcommand{\vb}{\vec{\beta}}
\newcommand{\vy}{\vec{v_y}}

\title{Collaborative Regression}
\author{Samuel M. Gross\thanks{email:smgross@stanford.edu} \and Robert
  Tibshirani\thanks{email:tibs@stanford.edu, Supported by NSF Grant DMS-99-71405 and National Institutes of Health Contract
N01-HV-28183}}
\date{\normalsize Departments of Statistics, and Health Research \& Policy, Stanford University}

\begin{document}

\maketitle

\begin{abstract}
{We consider the scenario where  one observes an outcome variable
and sets of features from  multiple assays, all measured  on the same set of samples.
  One approach that has been proposed for dealing with this type of data
 is ``sparse multiple canonical correlation
analysis'' (sparse mCCA).  All of the current sparse mCCA techniques 
are biconvex and thus have no guarantees about reaching
a global optimum.  We propose a method for performing sparse
supervised canonical correlation analysis (sparse sCCA), a specific
case of sparse mCCA when one of the datasets is a vector.  Our proposal for
 sparse sCCA is convex and thus does not face the same difficulties
as the other methods.
We derive efficient algorithms for this problem, and illustrate their use
on simulated and real data.}
\end{abstract}

\section{Introduction}
\label{sec:introduction}

The problem of combining data from multiple assays is an important
topic in modern biostatistics.  For many studies, the researchers have
more data than they know how to handle.  For example, a
researcher studying cancer outcomes may have both gene expression and
copy number data for a set of patients.  Should that researcher use
both types of predictors in their analysis?  Should any care be given
to distinguish the fact that these predictors are coming from
different assays and may have differing meanings?  If the researcher
needs to make future predictions based on only gene expression, is
there a way that having copy number data in a training set can help
those future predictions?  All of these are important questions that
are still up for debate.  

In this paper we propose a method for this problem called ``Collaborative
Regression'', a form of sparse
supervised canonical correlation analysis.
In Section \ref{sec:coll-regr} we define Collaborative
Regression (CollRe) and characterize its solution.  This involves explicit
closed form solutions for the unpenalized algorithm, as well as a
discussion of some useful convex penalties that can be applied.  Then, in Section
\ref{sec:coaching-variables} we explore the possibility of using
CollRe in a prediction framework.  While this may seem like an
intuitive use case, simulations suggest that CollRe is not able to
improve prediction error even over methods that do not take advantage
of the secondary dataset.

We look at using CollRe in a sparse sCCA framework in Section
\ref{sec:mult-canon-corr}, including a simulation study where we
compare CollRe to one of the leading competitors.  We
show how the penalized version can be applied to a real
biological dataset in Section \ref{sec:real-data-example}.  Finally,
in section \ref{appen} we explore how to efficiently solve the
convex optimization problem given by the penalized form of the algorithm.

\section{Collaborative Regression}
\label{sec:coll-regr}

Collaborative Regression is a tool designed for the scenario where 
there are groups of covariates that can
be naturally partitioned and a response variable.  Let us assume that we have
observed $n$ instances of $p_x+p_z$ covariates and a response.  We can
partition the covariates into two matrices, $X$ and $Z$, that are
$n\times p_x$ and $n\times p_z$ respectively.  The response values
are stored in a vector, $\y$, of length $n$.  Then Collaborative
Regression finds the $\hat \tx$ and $\hat \tz$ that minimize the
following objective function:

\begin{equation}
\label{crobj}
J(\theta_x, \theta_z)=\frac{b_{xy}}{2} \|\vec{y} - X\vec{\theta_x}\|^2 + \frac{b_{zy}}{2} \|\vec{y} - Z\vec{\theta_z}\|^2 + \frac{b_{xz}}{2}
\|X\vec{\theta_x} - Z\vec{\theta_z}\|^2
\end{equation}

This objective function seems natural for the multiple
dataset situation.  Basically, it says that we want to make
predictions of $\y$ based on $X$ or $Z$, but we will penalize
ourselves based on how different the predictions are.  Essentially,
the goal is to uncover a signal that is common to $X$, $Z$, and $\y$.

Consider trying to maximize the objective function (\ref{crobj}).  It is
easy to show using calculus that the optimal solution, $\hat\tx$ and
$\hat\tz$ will satisfy the following First Order Conditions:

\be\hat\tx = \frac{1}{\bxy+\bxz}(X^TX)^{-1}X^T(\bxy\y + \bxz
Z\hat\tz)
\ee
\be
\hat\tz = \frac{1}{\bzy+\bxz}(Z^TZ)^{-1}Z^T(\bzy\y + \bxz
X\hat\tx).
\ee

By substituting for $\hat\tz$ and solving, we can find a closed form
solution for $\hat\tx$:

\begin{multline}
\hat\tx = \left(I - \frac{\bxz^2}{(\bxy + \bxz)(\bzy +
    \bxz)}(X^TX)^{-1}X^TZ(Z^TZ)^{-1}Z^TX\right)^{-1}\\\left(\frac{\bxy}{\bxy
    + \bxz}(X^TX)^{-1}X^T\y + \frac{\bxy\bzy}{(\bxy
    + \bxz)(\bzy + \bxz)}(X^TX)^{-1}X^TZ(Z^TZ)^{-1}Z^T\y\right)
\end{multline}

In the above we have assumed that $X^TX$
and $Z^TZ$ are non-singular.  Assuming they are, and none of
the parameters are zero, then that
guarantees the invertibility of \newline
$\left(I - \frac{\bxz^2}{(\bxy + \bxz)(\bzy +
    \bxz)}(X^TX)^{-1}X^TZ(Z^TZ)^{-1}Z^TX\right)$.  Note that $X^TX$
and $Z^TZ$ will always be nonsingular in the classical case where
$\max(p_x,p_z) < n$.

\subsection{Infinite Series Solution}
\label{sec:inif-seri-solut}

Another way to characterize the optimal solution to the objective
function (\ref{crobj}) is as an infinite series.  Instead of solving for
$\hat\tx$ after substituting, consider instead what would happen if we
just continued substituting for $\hat\tx$ or $\hat\tz$ on the RHS.  Then, we
get an infinite series representation of $\hat\tx$.  Let $P_X =
X(X^TX)^{-1}X^T$ be the matrix that performs orthogonal projection
onto the column space of $X$ (and let $P_Z$ be defined similarly).
Then we can also write $\hat\tx$ as:

\begin{multline}
\hat\tx = \frac{\bxy}{\bxy + \bxz}(X^TX)^{-1}X^T\y + \frac{\bxz}{\bxy
  + \bxz}\frac{\bzy}{\bzy + \bxz}(X^TX)^{-1}X^TP_Z\y \\
+ \frac{\bxz}{\bxy + \bxz}\frac{\bxz}{\bzy + \bxz}\frac{\bxy}{\bxy +
  \bxz}(X^TX)^{-1}X^TP_ZP_X\y\\
+ \frac{\bxz}{\bxy + \bxz}\frac{\bxz}{\bzy + \bxz}\frac{\bxz}{\bxy +
  \bxz}\frac{\bzy}{\bzy + \bxz}(X^TX)^{-1}X^TP_ZP_XP_Z\y \dots
\end{multline}

If we let 
\begin{align*}
w_i &=&
\begin{cases}
\frac{\bxy}{\bxy + \bxz}\left(\frac{\bxz}{\bxy +
    \bxz}\frac{\bxz}{\bzy + \bxz}\right)^i & \text{if }i\text{ is even} \\
\frac{\bxz}{\bxy +
    \bxz}\frac{\bzy}{\bzy + \bxz}\left(\frac{\bxz}{\bxy +
    \bxz}\frac{\bxz}{\bzy + \bxz}\right)^i & \text{if }i\text{ is odd}
\end{cases}\\
\y^{(i)} &=&
\begin{cases}
(P_ZP_X)^i\y  & \text{if }i\text{ is even} \\
(P_ZP_X)^iP_Z\y & \text{if }i\text{ is odd}
\end{cases}
\end{align*}
Then
\be
\hat\tx = (X^TX)^{-1}X^T\displaystyle\sum_{i=0}^\infty w_i\y^{(i)}
\ee

Looking at the infinite expansion can help build some understanding of
what CollRe actually does.  We note that $\sum w_i = 1$, so
essentially CollRe is equivalent to regressing $X$ on the weighted
average of the $\y^{(i)}$'s.  Those $\y^{(i)}$'s trace out the path of
successive projections onto the column space of $X$ and $Z$.  As the
column spaces of $X$
and $Z$ are affine, it is known from Projection onto Convex sets that
the sequence will converge to the projection of $\y$ onto the
intersection of those two spaces.  In the case where the columns of
$X$ and $Z$ are linearly independent, $\y^{(i)}$ will eventually converge to
0.  Thus, CollRe is basically shrinking $\y$ towards the part that can
be explained by both $X$ and $Z$.

Additionally, we get some picture as to how the parameters $\{\bxy, \bzy, \bxz\}$ affect the solution.  $\frac{\bxy}{\bxy +
    \bxz}$ acts in large part to control the amount of shrinkage
  imposed on $\hat\tx$, while $\frac{\bzy}{\bzy + \bxz}$ does the same
  for $\hat\tz$.

\subsection{Penalized Collaborative Regression}
\label{sec:penal-coll-regr}

One nice aspect of the objective function (\ref{crobj}) is
that it is convex.  This means that the problem can still be easily
solved through convex optimization techniques if we add convex penalty
functions to the objective.  Thus, we can define Penalized
Collaborative Regression (pCollRe) as finding the minimizer of the
following objective:

\be
\label{pcrobj}
F(\theta_x, \theta_z)=\frac{b_{xy}}{2} \|\vec{y} - X\vec{\theta_x}\|^2 + \frac{b_{zy}}{2} \|\vec{y} - Z\vec{\theta_z}\|^2 + \frac{b_{xz}}{2}
\|X\vec{\theta_x} - Z\vec{\theta_z}\|^2 + P^x(\tx) + P^z(\tz)
\ee
where $P^x(\tx)$ and $P^z(\tz)$ are convex penalty functions.  Note
that some of the convex penalties that may warrant use include:
\begin{itemize}
\item The Lasso: $P^x(\tx)$ is an $\ell_1$ penalty on $\tx$, namely
  $P^x(\tx) = \lambda_x\|\tx\|_1$.  The lasso penalty is known to
  introduce sparsity into $\tx$ for sufficiently high values of $\lambda_x$.
\item Ridge: $P^x(\tx)$ is a (squared $\ell_2$ penalty on $\tx$, namely
  $P^x(\tx) = \lambda_x\|\tx\|^2_2$.  Ridge penalties help to smooth
  the estimate of $X^TX$ to ensure non-singularity.  This can be
  especially important in the high dimensional case where $X^TX$ is
  known to be singular.
\item The Fused Lasso: $P^x(\tx) = \displaystyle\sum_{i=2}^{i=p_x}\lambda_x|(\tx)_i
  - (\tx)_{i-1}|$.  The fused lasso will help to ensure that $\tx$ is
  smooth.  This can be helpful if there is reason to believe that the
  predictors can be sorted in a meaningful manner (as with copy number
  data).
\end{itemize}

In addition to the convex penalties above, situations may also call
for linear combinations of those penalties.  For example, the lasso
and ridge penalties are often combined to find sparse coefficients for
predictors that are highly correlated.  The lasso and fused lasso are
often combined to find sparse and smooth coefficient vectors.  In
Section \ref{appen} we discuss solving pCollRe efficiently in the case where the
penalty terms are asso penalties.  

\section{Using CollRe for Prediction}
\label{sec:coaching-variables}

One potentially appealing use of CollRe where we
want to make predictions of $\y$ for future cases where you will only
have the variables in $X$ available, and  $Z$ is only be available for a
training set.  Can the information
contained in $Z$ be used to help identify the correct direction in
$X$?  There are many practical situations in which this framework might be useful.  For example, maybe it is much
more costly to gather data with a lower amount of noise.
Alternatively, it could be that some data is not accessible until
after the fact; autopsy results may be very helpful in identifying
different types of brain tumors, but it is hard to use that
information to help current patients.

CollRe seems like it provides a natural way in which to perform
a regression with additional variables present only in the training
set.  Basically, it is saying that we want our future predictions to
agree with what we would have predicted given $Z$.  In this framework, CollRe is similar to ``preconditioning'' as
defined by Paul and others (2008) \cite{PBHT2006}.  Instead of
preconditioning on $Z$ and then fitting the regression, we are
simultaneously doing the preconditioning and fitting.

Looking at the infinite
series solution in Section \ref{sec:inif-seri-solut} it is clear that
performing CollRe is similar to doing ordinary regression after
shrinking $y$.  If that shrinkage on $y$ is done in such a way that it
reduces noise, we may ultimately expect ourselves to do better in
estimating the correct $\hat\tx$. We investigate this next.

\subsection{Simulated Factor Model Example}
\label{sec:simul-fact-model}

We decided to generate data from a factor model to test CollRe.  A
factor model seems natural for this problem,
and is a simply way to create  correlations between
$X$, $Z$, and $\y$.  Another reason the factor model was appealing is
because it is relatively easy to analyze, and given $\hat\tx$ and
$\hat\tz$ it is easy to compute statistics like the expected
prediction error or the correlations between linear combinations of
the variables.  More concretely, given values for parameters $n, p_x,
p_z, p_u, s_u, s_x, s_z,\text{ and } s_y$, we generate data according to the
following method:
\begin{enumerate}
  \item $\vy \in \mcalR^{p_u}$ distributed MVN(0,$I_{p_u}$)
  \item $\vec{v^x_j} \in \mcalR^{p_x}$ distributed iid
    MVN(0,$I_{p_x}$) for $j = 1,\dots,p_u$
  \item $V_x =  [\vec{v^x_1},\dots,\vec{v^x_{p_u}}]$
  \item $\vec{v^z_j} \in \mcalR^{p_z}$ distributed iid
    MVN(0,$I_{p_z}$) for $j = 1,\dots,p_u$
  \item $V_z =  [\vec{v^z_1},\dots,\vec{v^z_{p_u}}]$
  \item For $i = 1,\dots,n$:
    \begin{enumerate}
    \item $\vec{u_i} \in \mcalR^{p_u}$ distributed iid
      MVN(0,$s^2_uI_{p_u}$)
    \item $ y_i = \vy^T\vec{u_i} + \epsilon^y_i$ with $\epsilon^y_i$
    distributed N(0,$s^2_y$)
    \item $\vec{x_i} = V_x\vec{u_i} + \vec{\epsilon_x^i}$ with
      $\vec{\epsilon_x^i}$ distributed MVN(0,$s^2_xI_{p_x}$)
    \item $\vec{z_i} = V_z\vec{u_i} + \vec{\epsilon_z^i}$ with
      $\vec{\epsilon_z^i}$ distributed MVN(0,$s^2_zI_{p_z}$)
    \end{enumerate}
  \item $X = [\vec{x_1}, \dots,\vec{x_n}]^T, Z = [\vec{z_1},
    \dots,\vec{z_n}]^T, \text{ and } \y = [y_1,\dots,\y_n]^T$
\end{enumerate}

Thus, steps 1-5 generate the factors ($V = [V_X; V_Z; \vy]$) and step
6 generates the loadings ($u_i$) and noise.

In order to test the performance of CollRe in doing prediction, we
generated a set of factors from the above model with $n = 50,
p_x = p_z = 10, p_u = 3, s_u = s_x = s_z = s_y = 1$.  Then, for each
of 80 repetitions, we generated loadings and noise before fitting a
range of models.  CollRe was fit with $\bxy = \bzy = 1$ and a
variety of values of $\bxz$.  Additionally, at each level of $\bxz$ we
fit models with a range of ridge penalties.  Ridge Regression models were also fit,
which corresponds to $\bxz = 0$.  To evaluate the success of the fits,
we looked at prediction error based on using just $X$ relative to Ordinary Regression as well
as the sum correlation.  Here, by sum correlation, we mean
$\text{cor}(\vec{x_*}^T\tx,y_*) + \text{cor}(\vec{z_*}^T\tz,y_*) +
\text{cor}(\vec{x_*}^T\tx,\vec{z_*}^T\tz)$, where $(\vec{x_*}, \vec{z_*}, y_*)$
is a future observation (corresponding to making another pass through
step 6).

\begin{figure}[]
  \label{lowdim}
  \centering
    \includegraphics[width=\textwidth]{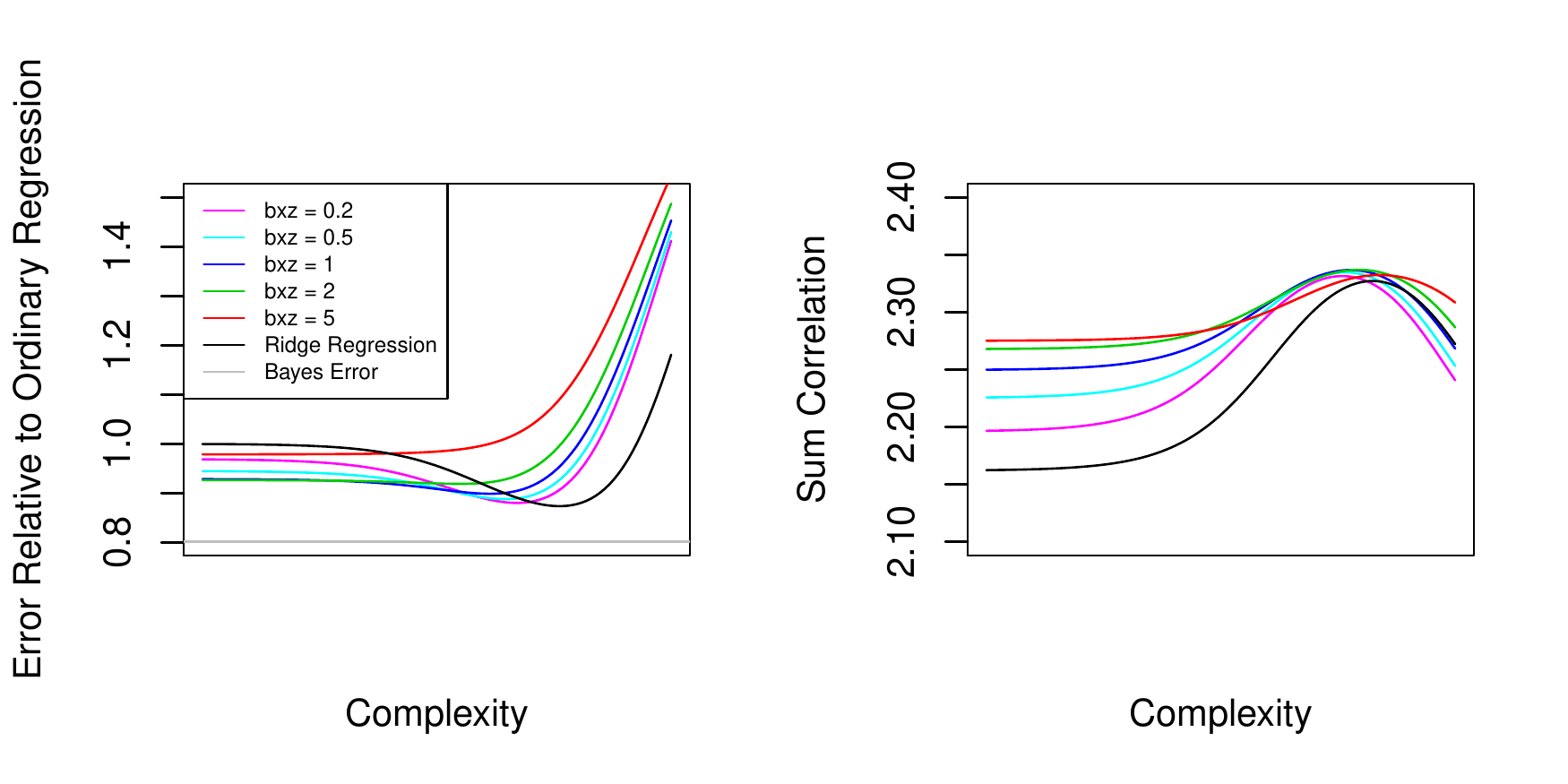} % This was generated by upenalized4
  \caption{\em Results of a simulation study to test the effectiveness of
    CollRe with $\ell_2$ penalty in a prediction framework.  Here, the points all the way to the left correspond to
    no $\ell_2$ penalty and the $\ell_2$ penalty increases (simpler models) as we
    move right along the x-axis.  The first plot shows prediction
    error for making future predictions based on $X$ only.  The second
    plot shows the theoretical sum correlation.  Values have been
    averaged over 80 repetitions.  As we can see, while CollRe
    outperforms ordinary regression with no penalty in terms of
    prediction error (the far left of
    the first plot), Ridge Regression achieves a lower minimum.  The
    second plot helps illuminate the reason; CollRe does a better job
    of maximizing the sum correlation, so it is sacrificing some of
    the correlation between $\y$ and $X\hat\tx$ in order to get a
    larger correlation between $X\hat\tx$  and $Z\hat\tz$}
\end{figure}

The results of the simulation, in Figure \ref{lowdim}, sheds some
light on the effectiveness of using CollRe to improve a regression of
$\y$ on $X$.   First, we note that ridge regression outperforms CollRe
at any choice of $\bxy$ and $\ell_2$ penalty for this particular problem.
At first, this might seem surprising given the fact that CollRe gets
the advantage of using $Z$ and ridge regression does not.  When
looking at the sum correlation though, we see that CollRe
outperforms ridge.  This suggests that the reason CollRe is doing worse
on predicting $\y$ is because it is focusing on the distance between
$X\hat\tx$ and $Z\hat\tz$ instead of just the typical RSS.
Essentially, CollRe is giving up a little of the fits involving $\y$
in order to get a higher correlation between $X\hat\tx$ and $Z\hat\tz$.
It seems that CollRe is more naturally suited for
supervised canonical correlation analysis, discussed next.

\section{Supervised Canonical Correlation Analysis (sCCA)}
\label{sec:mult-canon-corr}

Canonical Correlation Analysis (CCA) is a data analysis technique that
dates back to Hotelling (1996) \cite{hotelling1936}. 
  Given two sets of centered variables, $X$ and
$Z$, the goal of CCA is to find linear combinations of $X$ and $Z$
that are maximally correlated.  Mathematically, CCA performs the
following constrained optimization problem:
\[(\hat\tx,\hat\tz) = \arg\max_{\tx,\tz} \tx^TX^TZ\tz \text{ such that
} \tx^TX^TX\tx \le 1, \tz^TZ^TZ\tz \le 1\]

In this form, it is possible to derive a closed form solution for
CCA using matrix decomposition techniques.  Namely, $\hat\tx$
will be the eigenvector corresponding to the largest eigenvalue of
$(X^TX)^{-1}X^TZ(Z^TZ)^{-1}Z^TX$.  A similar expression can be found
for $\hat\tz$ by switching the roles of $X$ and $Z$.

CCA might be a useful tool for finding a signal that is common
to both $X$ and $Z$, but there is no guarantee that the discovered signal will
also be associated with $\y$.  To approach this issue, a generalization
of CCA called Multiple Canonical Correlation Analysis (mCCA) was
developed.  mCCA allows for more than 2 datasets and seeks to find a
signal that is common to all of the datasets.  The case we have, where
the third dataset is a vector, can be thought of as a special case of
mCCA that we will call Supervised Canonical Correlation Analysis
(sCCA).

There are many techniques that approach the mCCA problem.  Most of
them focus on optimizing a function of the correlations between the
various datasets.  Gifi (1990) \cite{gifi1990nonlinear} provides an overview of many of the
suggestions that have been made for this problem.  One example of an
optimization problem that people would call mCCA is based on trying to
maximize the sum of the correlations:

\be
\label{mccaobj}
\{\vec{\theta_i}\}_{i = 1,\dots,k} = \arg\max \sum_{i < j} \vec{\theta_i}^TX_i^TX_j\vec{\theta_j} \text{ such that
} \vec{\theta_i}^TX_i^TX_i\vec{\theta_i} \le 1\text{  } \forall i
\ee

Now the optimization problem above is multiconvex as long as each of
the $X_i^TX_i$ are non-singular.  This means that a local
optimum can be found by iteratively maximizing over each $\theta_i$
given the current values of the rest of the coefficients.

\subsection{Sparse sCCA}
\label{sec:sparse-scca}

For high dimensional problems (where $p_i >>> n$ for at least one $i$), several issues emerge when doing sCCA.
First, the constraints given in equation (\ref{mccaobj}) are
no longer strictly convex constraints because $X_i^TX_i$ is necessarily
singular for at least one $i$.  This means that the problem cannot be
as easily solved by an iterative algorithm.

One approach that some people take to this problem is to add a ridge
penalty on the coefficients.  As with ridge regression, adding a ridge
penalty will effectively replace $X_i^TX_i$ with $X_i^TX_i + \lambda_iI$ ($I$ being the identity matrix), which
will then be non-singular.  This means that the mCCA problem in
equation (\ref{mccaobj}) can be solved by adding a ridge penalty.
Examples of works where people have pursued this method include
Leurgans and others (1993)
\cite{leurgans1993}.  Another approach is pursued by Witten and
Tibshirani (2009) \cite{WittenTibsSAGMB09} where $X_i^TX_i$ is
replaced by $I$ in order to ensure strict convexity of the constraints.

Now, even after adjusting to make sure that the constraints (or
penalties in the Lagrange form) are convex, there is still another
issue that the high dimensional regime adds.  For many problems in the
high dimensional regime, the goal of the problem is to do some sort of
variables selection.  After all, it is much more useful for a
biologist to uncover 30 genes or pathways that are particularly
important in a process than it is to uncover 30,000 coefficient values
that are all fairly noisy anyway.  Another way to state that is that
we want to find coefficients that are sparse (mostly 0).  There has
been a lot of work in sparse statistical methods following the
introduction of the lasso by Tibshirani (1996)
\cite{Ti96}.  
Witten and Tibshirani (2009) \cite{WittenTibsSAGMB09} offer the following
optimization problem to perform sparse mCCA:

\[\{\vec{\theta_i}\}_{i = 1,\dots,k} = \arg\max \sum_{i < j} \vec{\theta_i}^TX_i^TX_j\vec{\theta_j} \text{ such that
} \vec{\theta_i}^T\vec{\theta_i} \le 1, \|\vec{\theta_i}\| < c_i
\forall i\]
where the $c_i$ can be chosen to impose the desired level of sparsity
on each coefficient vector.  Note that further convex constraints (or penalties) can be
added to the above such as the fused lasso or non-negativity
constraints.  As with the other methods, this problem is multiconvex
and can be solved through an iterative algorithm.

Note however, that a multiconvex problem may be particularly hard to
solve in a high dimensional space.  While we know the algorithm will
converge to a local optimum, we would ideally like to find the global
optimum.  For a low dimensional space this can be mostly resolved by
doing multiple starts from random points in the coefficient space.
With enough starts we believe that we can search the space
sufficiently well that our best local optimum is at least close to
globally optimum.  This logic breaks down in high dimensional spaces
because it is impossible to sufficiently search the space without
exponentially many starting points.  This means that while the above
methods for Sparse mCCA have outputs, we won't know whether those
outputs are even optimizing the criteria in high dimensions.  

\begin{figure}
\label{vsdanielaobj}
\centering
\includegraphics[width=.48\textwidth]{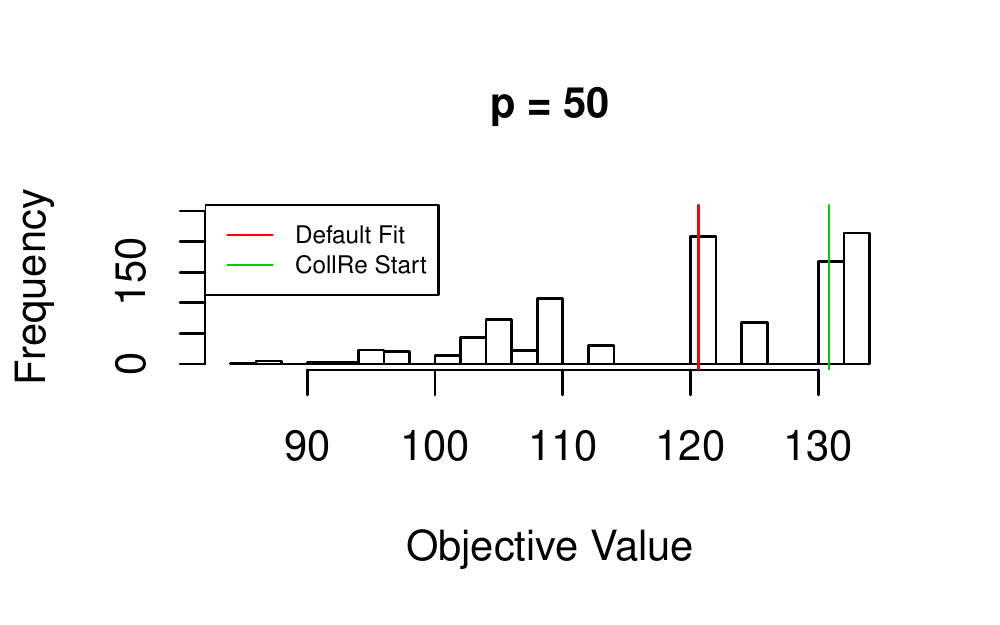}
\includegraphics[width=.48\textwidth]{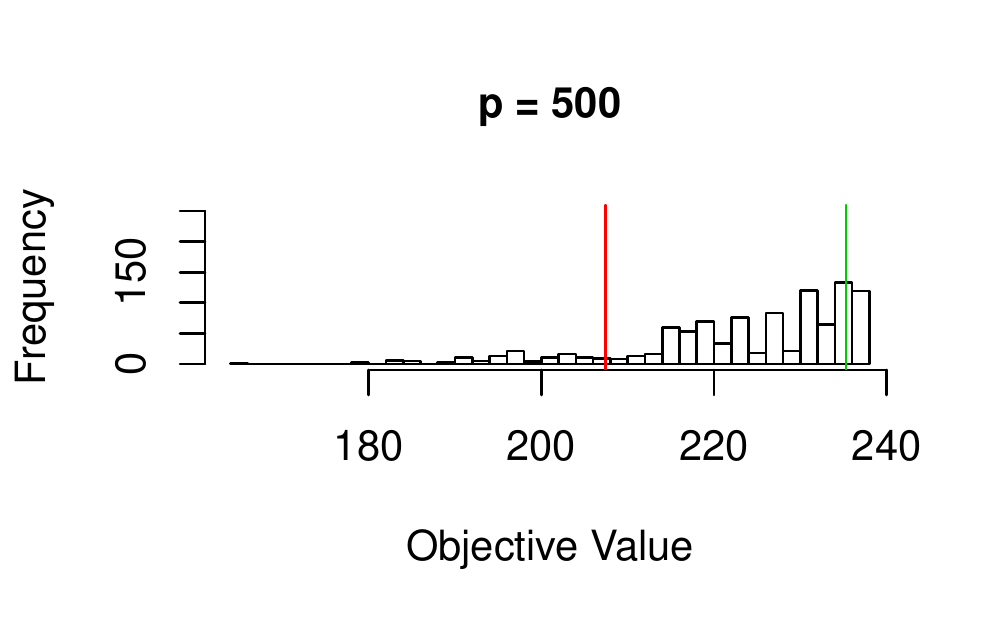}\\
\includegraphics[width=.48\textwidth]{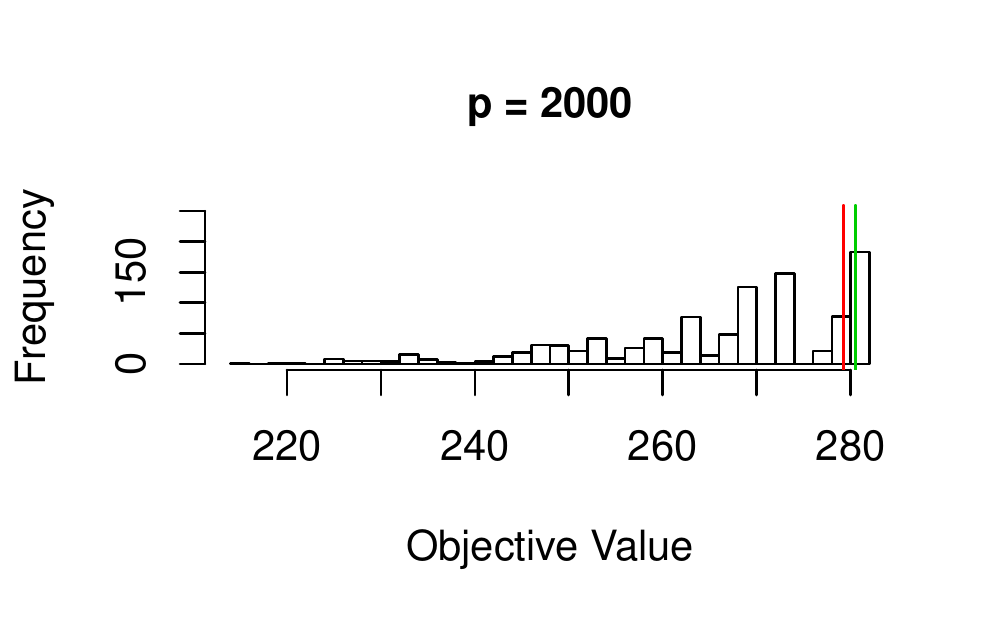}
\includegraphics[width=.48\textwidth]{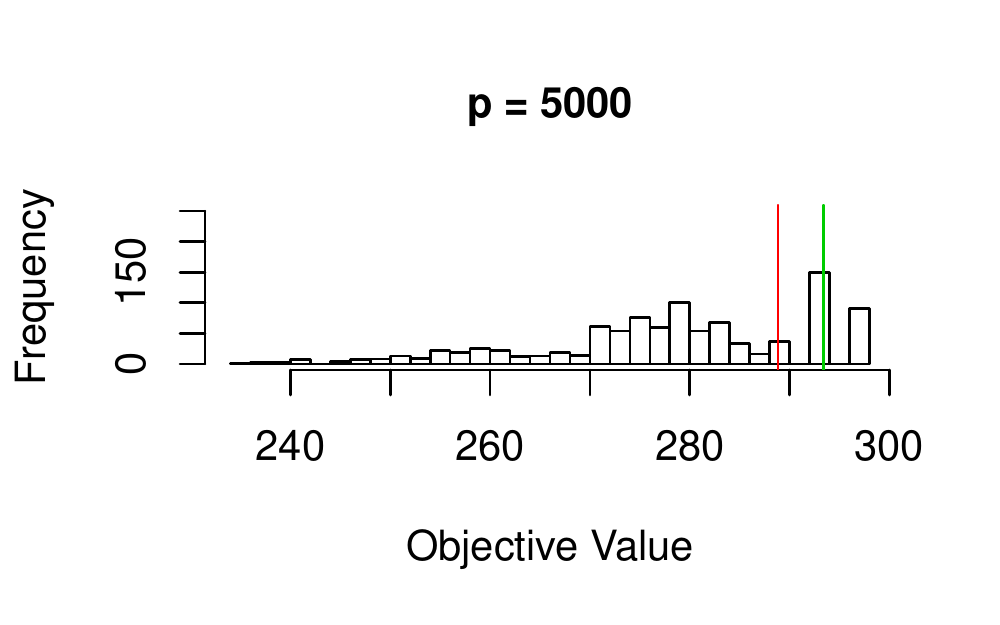}
  \caption{\em Result of a simulation to see how close the locally
   optimal solutions to MultiCCA end up to the global optimum.  Histograms of objective values of MultiCCA from $1000$ random
   starts.  As we can see, the random starts end up at a variety of
   local optima, and using the results of CollRe as a starting point
   often outperforms the default start which is based on an singular
   value decomposition.  In each case, $n = 50$.}
\end{figure}

We generated some data to test the extent to which MultiCCA gets
caught in local optima.  These datasets have $n = 50, p_u = 30, s_u =
\sqrt{1/10}, s_x = s_z = s_y = 1$.  For $p = 50,500,2000, \text{ and }5000$, we
generated a dataset with $p_x = p_z = p$ and then ran
MultiCCA from 1000 random (uniform on the unit sphere) start locations.  Figure \ref{vsdanielaobj}
shows histograms of the resulting objective values.  The vertical
lines correspond to the default starting point of MultiCCA, and a
starting point that is based on a penalized CollRe solution.  As we
can see, there are many local optima that emerge especially in higher
dimensions.  One interesting thing is that the CollRe starts typically
end up in a better solution than the default starts provided by the
MultiCCA function.

Another option to perform sparse high-dimensional sCCA was suggested
by Witten and Tibshirani (2009) \cite{WittenTibsSAGMB09}.  She
suggests that a method of supervision similar to Bair and others
(2006) \cite{BHPT2006}
can be used: before doing a fit, all of the variables are
screened against $\y$.  Only the ones that have correlation above some
threshold will be passed along to a CCA model.  This method
can also be used to add a supervised component to any of the methods
that can be used to perform CCA.  The main issue with this approach is that it does the
supervision in a way that is completely univariate.

\subsection{Penalized Collaborative Regression as Sparse sCCA}
\label{sec:coll-regr-as}

Consider one of the three terms from our objective function:

\be
\label{eq:myterm}
\min \|X\tx - Z\tz\|^2 = \tx^TX^TX\tx + \tz^TZ^TZ\tz - 2\tx^TX^TZ\tz
\ee

Now let's compare that to the following version of the CCA objective:

\[\min -\tx^TX^TZ\tz\ \text{ such that } \tx^TX^TX\tx \le 1,
\tz^TZ^TZ\tz \le 1\]

We can convert the CCA problem from its bounded form into the
Lagrange form as follows:

\[\min -\tx^TX^TZ\tz + \lambda_x \tx^TX^TX\tx + \lambda_z
\tz^TZ^TZ\tz\],
where $\lambda_x$ and $\lambda_z$ are chosen appropriately to enforce
the unit variance constraint.  In this way CCA can also be characterized as
a penalized optimization problem.  The difference between the term from our objective, and the penalized
form of CCA is that instead of using $\lambda_x$ and $\lambda_z$ in
order to enforce unit variance, we choose the values that would result
in the objective being convex instead of merely biconvex.  

Now it is worth noting that an unenviable fact about the
penalty used in equation (\ref{eq:myterm}) is that it results in the
minimum being achieved by setting all of the coefficients equal to
zero.  Fortunately, CollRe avoids this issue because  the two terms
involving $\y$.  

Thus, CollRe with $\bxy = \bzy = \bxz = 1$ is very similar to doing a
sum of correlations mCCA as in the equation (\ref{mccaobj}), with the
exception that we have picked the penalties that allow for convexity
instead of the penalties that correspond to unit variance.

As discussed in Section \ref{sec:penal-coll-regr} one of the
advantages of CollRe is the simplicity with which convex penalties can
be added to the objective function.  Thus, it is easy to convert
CollRe into a form that is appropriate for sparse sCCA by adding
penalties just as in Witten and Tibshirani (2009) \cite{WittenTibsSAGMB09}.

\begin{figure}[]
  \label{vsdaniela}
  \centering
    \includegraphics[width=\textwidth]{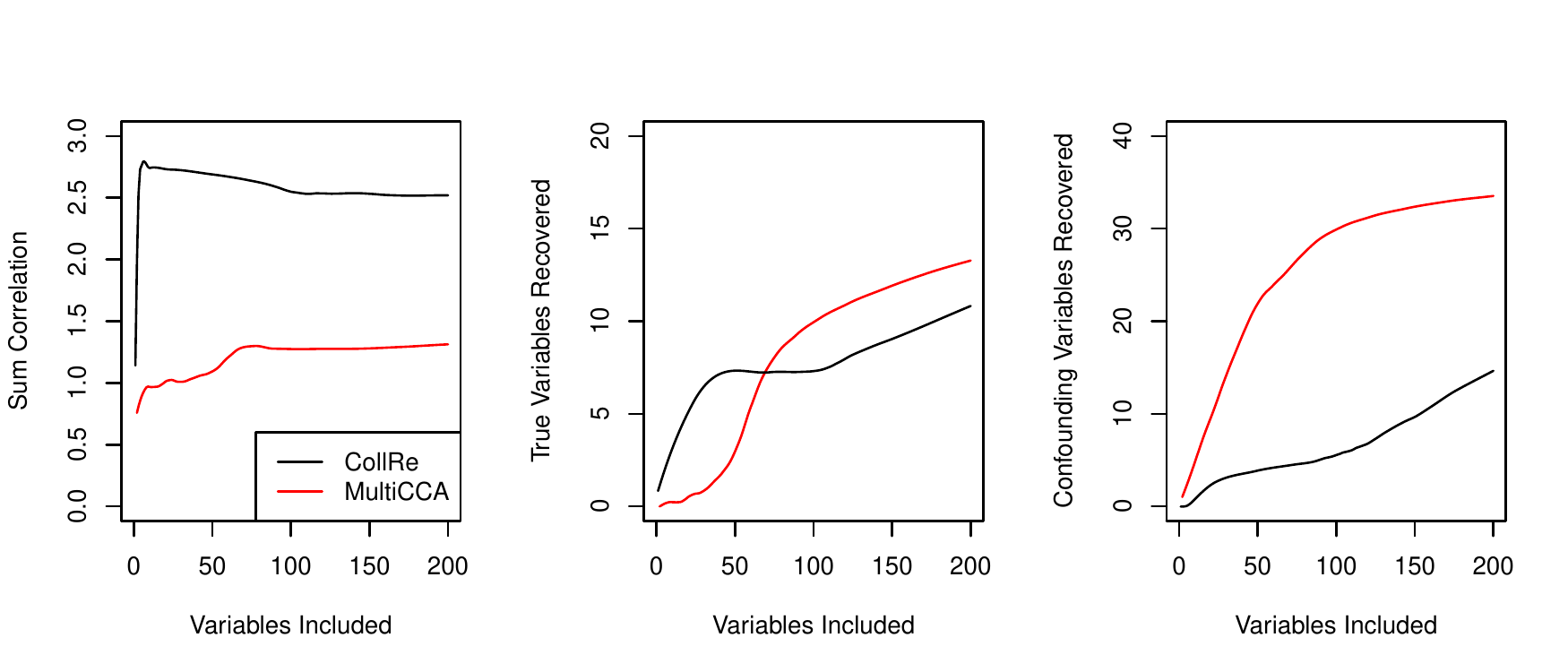}
    % combination of dec13pen5.R on solomon and dec13_penalized.R in
    % dropbox
   % The current version (highdimvsdanielawithconpu) has issues b/c I
   % made the sd on the confounders huge.
  \caption{\em Results of a simulation to compare CollRe and MultiCCA
    in performing sparse supervised mCCA.  For each repetition, a
    dataset with $n = 50, p_x = p_z = 500$ is created.  For both $X$
    and $Z$, $40$ of the predictors are confounding variables and $20$
    of the predictors are true variables (the rest are null).  Confounding
    variables are the ones that share a signal between $X$ and $Z$, but
    not $\y$.  The true variables share a signal between all three
    datasets.  Values have been averaged over $80$ repetitions.
    MultiCCA is much more susceptible to picking up the confounding
    variables, and thus has a much harder time achieving high
    correlations.  Interestingly, while CollRe finds many more true
    variables at first, after $70$ or so included variables MultiCCA
    starts finding more.}
\end{figure}

To compare CollRe against a competing algorithm for sparse sCCA, we generated data from the above model with $n = 50,
p_x = p_z = 20, p_u = 3, s_u = 1, s_x = s_z = s_y = 1/2$.  Then, we
added $40$ variables to both $X$ and $Z$ that were generated from $3$
new factors that have no effect of $\y$.  These $40$ variables act as
confounding variables that reflect an effect we do not want to
uncover.  This could correspond to a batch effect in the measurements,
or maybe some other underlying difference among the sampled patients.
Finally, we
added another $440$ columns to $X$ and $Z$ that were just independent
gaussians to act as null predictors.  We ran both CollRe with a lasso
penalty and Wittens MultiCCA from the PMA package with an 
 $\ell_1$
constraint each over a range of parameter values.  This process was
repeated 80 times with new loadings and noise each time but the same
factors.  Figure \ref{vsdaniela} shows the average (over repetitions) theoretical sum correlation
for future observations, as well as the recovery of true predictors,
against a range of nonzero coefficients that corresponds to a range of
penalty parameters.

From the results, we can see that CollRe does a much better job of
finding coefficients that have high sum correlation.  MultiCCA seems
to get caught in the trap set by the confounding variables, which makes it
harder to raise the sum correlation much above 1 (a perfect
correlation between $\vec{x_*}^T\tx$ and $\vec{z_*}^T\tz$ with no
relation to $y_*$).  Interestingly,
while MultiCCA does worse than CollRe on recovery of true variables
for the first 70 or so variables added, it seems to do a better job of
recovering true variables after that point.  It is unclear what
exactly is causing that transition in this problem.

\section{Real Data Example}
\label{sec:real-data-example}

To demonstrate the applicability of penalized CollRe, we also ran it on a high
dimensional biological dataset.  We used a neoadjuvant breast cancer
dataset that was provided by our collaborators in the Division of
Oncology at the Stanford University School of Medicine.  Details about
the origins of the data can be found at ClinicalTrials.gov using the
identifier NCT00813956.  This dataset consists of $n = 74$
patients who underwent a particular breast cancer treatment.  Before
treatment, the patients had measurements taken on their gene expression as well
as copy number variation.  In all, after some pre-processing, there were $p_x = 54,675$ gene expression
measurements per patient and $p_z = 20349$ copy number variation
measurements.  Additionally, each patient was given a RCB score six
months after treatment that corresponds to how effective the treatment
was.  The RCB score is essentially a composite of various metrics on
the tumor: primary tumor bed area, overall \% cellularity, diameter of
largest axillary metastasis, etc.

The goal of the analysis is to select a set of gene expression
measurements that are highly correlated with a particular pattern of copy number
variation gains or losses.  That said, we are only interested in sets
that also correlate with the RCB value.  As such, it is the perfect
opportunity to employ CollRe.

Due to computational limitations and issues with noise in the
underlying measurements, some further pre-processing was done to the data.
First, the gene expression measurements were screened by their
variance across the subjects.  Only the top $28835$ gene expression
genes were kept.  For the copy number variation measurements we needed
to account for the fact that for each patient the copy number
variation measurements are VERY highly autocorrelated because they had
already been run through a circular binary segmentation algorithm (a
change point algorithm used to smooth copy number variation data).  We use a fused lasso penalty to help correct for the
fact that we don't really have gene level measurements.  However,
doing fused lasso solves can be very slow for large $p$, so we took
consecutive triples of the copy number variation measurements and
averaged them.  This reduced the number of copy number variation
measurements to $6783$.  Our new $X$ and $Z$ matrices were scaled and
centered, and then CollRe on the dataset with
$\bxy = \bxz = \bzy = 1$ and the following parameters and penalty terms:
\[P^x(\tx) = \lambda_x(.9\|\tx\|_1 + .1\frac{1}{2}\|\tx\|_2^2)\]
\[P^z(\tz) = 4\|\tz\|_1 + 200 \displaystyle\sum_{i=2}^{i=p_z}|(\tz)_i
  - (\tz)_{i-1}|\]

\begin{figure}[]
  \label{fig:realex}
  \centering
    \includegraphics[width=\textwidth]{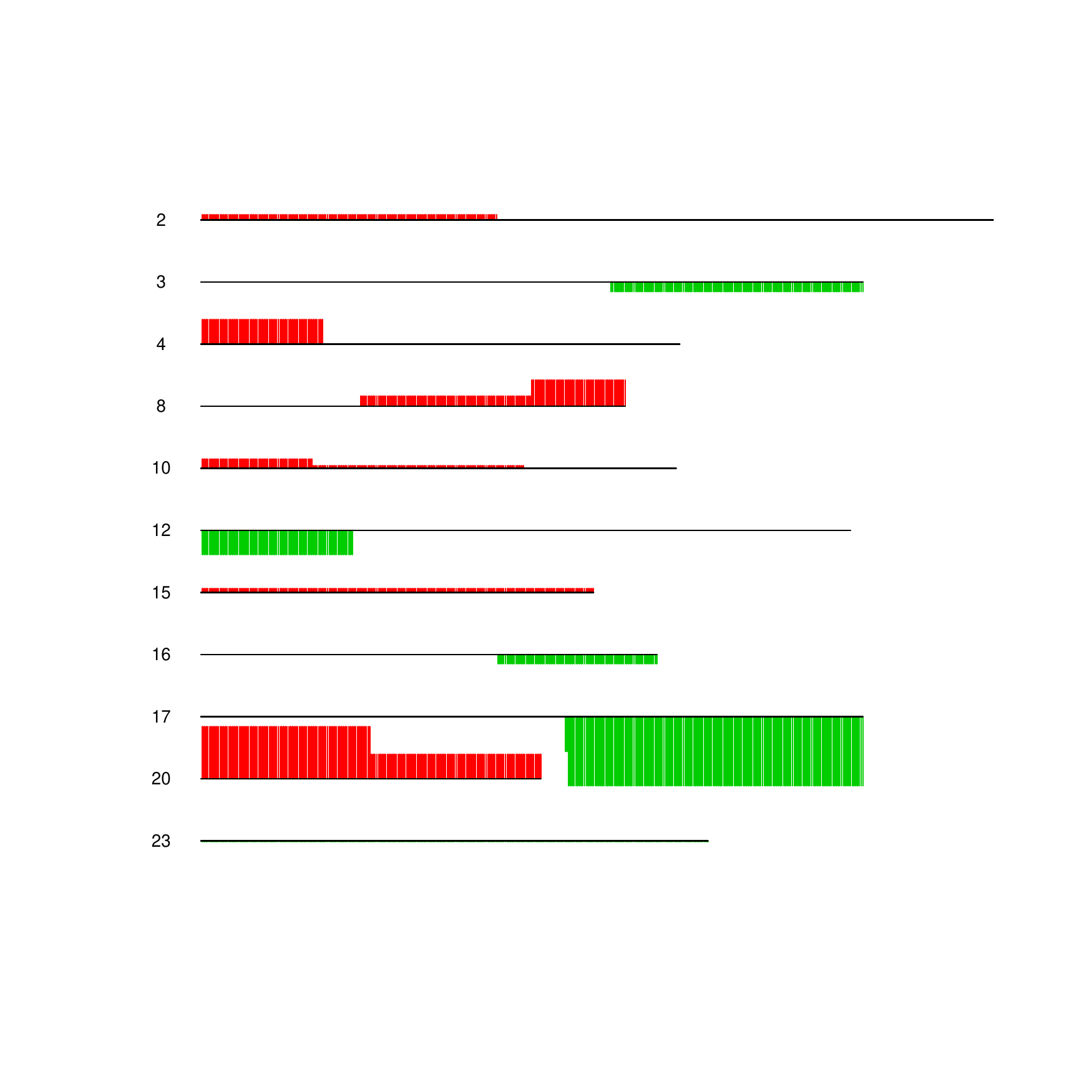}
   \caption{\em The resulting vector of coefficients for the copy
    number variation data from running CollRe on the RCB dataset.
    Regions with
    positive coefficients (amplification associated with higher RCB)
    are darker and appear above the line.  Regions with negative
    coefficients are lighter and appear below the line.  The size of
    the bars are proportional to the coefficient values.  Missing
    chromosomes had no nonzero coefficients.  The piece-wise constant
    nature of the 
    coefficient vector is due to the use of a fused lasso.}
\end{figure}

We searched a grid of $\lambda_x$ in order to find a
solution with about 50 nonzero coefficients in each set of variables.  This
corresponds roughly with the number of genes a collaborator thought
she would be able to reasonably examine for plausible connections.
The penalty terms on $\tz$ were chosen in a way that the selected
coefficients looked reasonably smooth.  The resulting $\hat\tz$ vector
can be seen in Figure \ref{fig:realex}

\section{Solving CollRe with Penalties}
\label{appen}

In Section \ref{sec:penal-coll-regr} we mentioned that CollRe is
solvable with a variety of penalty terms added.  In fact, due to the
nature of the CollRe objective, it can often be solved for common
penalty terms using out of the box penalized regression solvers.  To
make this concrete, let us focus on CollRe with the addition of $\ell_1$ penalties.

 Consider then, the objective function with penalty terms:
\begin{multline}
\label{fullobj}
J(\tx,\tz;X,Z,\y, \bxy,\bzy,\bxz,\lx,\lz) = 
\frac{\bxy}{2} \|\y - X\tx\|^2 + \frac{\bzy}{2} \|\y - Z\tz\|^2 + \frac{\bxz}{2}
\|X\tx - Z\tz\|^2 + \\\lx\|\tx\|_1 + \lz\|\tz\|_1 
\end{multline}

We note that (\ref{fullobj}) is a convex function, so we can optimize
it by iteratively optimizing over $\tx$ and $\tz$.  For a given value
of $\tz$, the optimal $\tx$ is given by:

\be
\hat\tx = \text{LASSO}(X,\vec{y^*},\frac{\lx}{\bxy+\bxz}) \text{ , where } \vec{y^*}
= \frac{\bxy}{\bxy+\bxz}\y + \frac{\bxz}{\bxy+\bxz}Z\tz
\ee

Here, LASSO$(\tilde X,\tilde{\vec{y}},\tilde\lambda)$ is
the solution to the $\ell_1$ penalized regression problem:
\be
\hat{\vec{\beta}} = \arg \min_{\vec{\beta}} \|\tilde{\vec{y}}-\tilde X\vec{\beta}\|^2 +
\tilde \lambda \|\vec{\beta}\|_1
\ee

An equivalent solution of $\hat\tz$ given $\tx$ can be found
by symmetry.  Thus, by iterating back and forth between these two
$\ell_1$ penalized regression problems, we are guaranteed to the optimum of equation
(\ref{fullobj}).  This means it is trivial to write a solver for
CollRe using $\ell_1$ penalties as long as you have access to a solver for
regression with $\ell_1$ penalties.  Many such functions can be found in R
packages, including the popular \textbf{glmnet} function in the self
titled package.

\subsection{Proof of Correctness of Algorithm (CollRe with $\ell_1$ Penalty)}
\label{sec:proof-algorithm}

Let $\tilde J$ be the LASSO  criterion:
\be
\tilde J(\tilde{\vec{\beta}}; \tilde X, \tilde{\vec{y}}, \tilde \lambda) =  \|\tilde{\vec{y}}-\tilde X\tilde{\vec{\beta}}\|^2 + \tilde\lambda \|\tilde{\vec{\beta}}\|_1
\ee

Then we see that $\tilde J$ has subgradient:
\be
\frac{\partial \tilde{J}}{\partial \tilde{\vec{\beta}}} =  \tilde X^T\tilde X\tilde{\vec{\beta}} - \tilde X^T\tilde{\y} + \tilde \lambda s(\tilde{\vec{\beta}})
\ee

Compare this to the subgradient of $J$ with respect to $\tx$:
\be
\label{subgJ}
\frac{\partial J}{\partial \tx} = (\bxy + \bxz)X^TX\tx - X^T(\bxy \y +
\bxz Z\tz) + \lx s(\tx)
\ee

Dividing (\ref{subgJ}) by $\bxy+\bxz$ and substituting $\vec{y^*} =
\frac{\bxy}{\bxy+\bxz}\y + \frac{\bxz}{\bxy+\bxz}Z\tz$ completes the
proof.

\subsection{Augmented Data Version}
\label{sec:augm-data-vers}

For some selections of penalties, parameters, and solvers, CollRe 
can be fit using an augmented data approach.  This means that the
solution can be found in just one call to a solver instead of having to
iterate.  In practice, this can increase the rate of convergence and
reduce total computation time.

Let us return to the example of trying to fit CollRe with the addition
of an $\ell_1$ penalty.  Consider the following LASSO problem:
\be
\tilde X = \left[ \begin{array}{cc}
\sqrt{\bxy} X & 0 \\
0 & \sqrt{\bzy} \frac{\lx}{\lz} Z \\
\sqrt{\bxz} X & -  \sqrt{\bxz} \frac{\lx}{\lz} Z \end{array}\right],
\tilde \y = \left[ \begin{array}{c}
\y \\
\sqrt{\bzy}\y \\
0 \end{array}\right],
\tilde \vb = \left[ \begin{array}{c}
\tx \\
\frac{\lz}{\lz}\tz \end{array}\right]
\ee

\be
\label{lassoaugver}
\hat{\tilde \vb} = \arg\min_{\tilde \vb} \|\tilde X \tilde{\vb} -
\tilde \y\|^2 + \lx\|\tilde{\vb}\|_1
\ee

It can be easily verified that equation (\ref{lassoaugver}) is exactly
the CollRe with $\ell_1$ penalty fit for the parameters given.  Essentially, this means that
instead of iterating between LASSO solves with $(\tilde{n} =
n,\tilde{p} = p_x)$ and $(\tilde{n} = n,\tilde{p} = p_z)$ until
convergence, we only do one solve with $(\tilde{n} =
3n,\tilde{p} = p_x + p_z)$.  Because we expect $n <<< \max (p_x,p_z)$,
we don't expect tripling $\tilde{n}$ to have much effect on run time.
Further, due to active set rules that are built into packages like the R package \textbf{glmnet}, even if we double $\tilde{p}$ it
should not have too large an effect on run time (Friedman and others
(2010) \cite{FHT2010}).  

We ran some simulations that involve generating $X,Z,$ and $\y$ from
independent standard normal draws.  we then fit CollRe with Elastic Net to the data
setting all of the parameters equal to one (except $\ltx = \ltz = 0$).
For $n = 100$, $p_x = p_z = 2000$ the normal version of CollRe with $\ell_1$ 
penalty ($\lx = \lz = 1$) takes about 2.2
seconds to run on a 2010 Macbook Pro.  The augmented version only takes 0.6
seconds to run.  The augmented version also achieves a lower value for
the objective function (8.127974 compared to 8.128410), so the speedup is
not just coming from a premature convergence.

\section{Discussion}
\label{sec:discussion}

In this paper, we introduced a new model called Collaborative
Regression, which can be used in settings where one has two sets of predictors
and a response variable for a set of observations.  We explored the
possibility of using CollRe in a prediction framework, but ultimately
decided that it was not particularly well suited for that task.

We then discussed the
problem of sparse supervised Canonical Correlation Analysis, which
seems to be an increasingly interesting problem for biostatistics.
While current approaches to sCCA are biconvex and don't
necessarily lend themselves to a sparse generalization, CollRe does
not suffer from those same issues.  We used several simulations and real data to
explore both the issues of biconvexity in high dimensions, as well as
the performance of CollRe.  

%TODO!!!
% funding section: Supported by NSF Grant DMS-99-71405 and National
% Institutes of Health Contract N01-HV-28183
% acknowledgements

\section*{Acknowledgments}
The authors thank S. Vinayak, M. Telli, and J. Ford for 
comments regarding the use of CollRe as well as providing the dataset used
in section \ref{sec:real-data-example}.  We also thank T. Hastie,
J. Taylor, D. Donoho, and D. Sun for comments regarding the development of the
CollRe algorithm.

%\bibliographystyle{agsm} Ryan's template used this
%\bibliographystyle{biorefs}
%\bibliography{../../mybib.bib}{}

\end{document}